\documentclass[aps,prl,amsmath,amssymb,reprint,superscriptaddress,showpacs]{revtex4-1}
\usepackage{graphicx}

\begin{document}

\title{Neutron-skin thickness from the study  of the anti-analog giant
  dipole resonance}
\author{A. Krasznahorkay}
\email[]{kraszna@atomki.hu}
\affiliation{Inst. of Nucl. Res. (ATOMKI), H-4001 Debrecen, P.O. Box 51, Hungary}

\author{L. Stuhl}
\affiliation{Inst. of Nucl. Res. (ATOMKI), H-4001 Debrecen, P.O. Box 51, Hungary}

\author{M. Csatl\'os}
\affiliation{Inst. of Nucl. Res. (ATOMKI), H-4001 Debrecen, P.O. Box 51, Hungary}

\author{A. Algora}
\affiliation{Inst. of Nucl. Res. (ATOMKI), H-4001 Debrecen, P.O. Box 51, Hungary}

\author{J. Guly\'as}
\affiliation{Inst. of Nucl. Res. (ATOMKI), H-4001 Debrecen, P.O. Box 51, Hungary}

\author{J. Tim\'ar }
\affiliation{Inst. of Nucl. Res. (ATOMKI), H-4001 Debrecen, P.O. Box 51, Hungary}

\author{N. Paar}
\affiliation{Physics Department, Faculty of Science, University of Zagreb,
  Croatia}
\author{D. Vretenar}
\affiliation{Physics Department, Faculty of Science, University of Zagreb,
  Croatia}

\author{K.~Boretzky}
\affiliation{GSI, Helmholtzzentrum f\"{u}r Schwerionenforschung GmbH,
      Darmstadt, Germany}
\author{M. Heil}
\affiliation{GSI, Helmholtzzentrum f\"{u}r Schwerionenforschung GmbH,
      Darmstadt, Germany}
\author{Yu.A. Litvinov}
\affiliation{GSI, Helmholtzzentrum f\"{u}r Schwerionenforschung GmbH,
      Darmstadt, Germany}
\author{D. Rossi}
\affiliation{GSI, Helmholtzzentrum f\"{u}r Schwerionenforschung GmbH,
      Darmstadt, Germany}
\author{C. Scheidenberger}
\affiliation{GSI, Helmholtzzentrum f\"{u}r Schwerionenforschung GmbH,
      Darmstadt, Germany}
\author{H. Simon}
\affiliation{GSI, Helmholtzzentrum f\"{u}r Schwerionenforschung GmbH,
      Darmstadt, Germany}
\author{H. Weick}
\affiliation{GSI, Helmholtzzentrum f\"{u}r Schwerionenforschung GmbH,
      Darmstadt, Germany}

\author{A.~Bracco}
\affiliation{INFN sez. Milano,Via Celoria 16, 20133 Milano, Italy}
\affiliation{Universita degli Studi di Milnao,Dipartimento di Fisica, Milano,Italy}
\author{S.~Brambilla}
\affiliation{INFN sez. Milano,Via Celoria 16, 20133 Milano, Italy}
\author{N. Blasi}
\affiliation{INFN sez. Milano,Via Celoria 16, 20133 Milano, Italy}
\author{F. Camera}
\affiliation{INFN sez. Milano,Via Celoria 16, 20133 Milano, Italy}
\affiliation{Universita degli Studi di Milnao,Dipartimento di Fisica, Milano,Italy}
\author{A. Giaz}
\affiliation{INFN sez. Milano,Via Celoria 16, 20133 Milano, Italy}
\affiliation{Universita degli Studi di Milnao,Dipartimento di Fisica, Milano,Italy}
\author{B. Million}
\affiliation{INFN sez. Milano,Via Celoria 16, 20133 Milano, Italy}
\author{L. Pellegri}
\affiliation{INFN sez. Milano,Via Celoria 16, 20133 Milano, Italy}
\affiliation{Universita degli Studi di Milnao,Dipartimento di Fisica, Milano,Italy}
\author{S. Riboldi}
\affiliation{INFN sez. Milano,Via Celoria 16, 20133 Milano, Italy}
\affiliation{Universita degli Studi di Milnao,Dipartimento di Fisica, Milano,Italy}
\author{O.~Wieland}
\affiliation{INFN sez. Milano,Via Celoria 16, 20133 Milano, Italy}

\author{S.~Altstadt}
\affiliation{Goethe-Universit\"at Frankfurt am Main, Germany}
\affiliation{GSI, Helmholtzzentrum f\"{u}r Schwerionenforschung GmbH,
      Darmstadt, Germany}
\author{M. Fonseca}
\affiliation{Goethe-Universit\"at Frankfurt am Main, Germany}
\author{J. Glorius}
\affiliation{Goethe-Universit\"at Frankfurt am Main, Germany}
\author{K. G\"obel}
\affiliation{Goethe-Universit\"at Frankfurt am Main, Germany}
\author{T. Heftrich}
\affiliation{Goethe-Universit\"at Frankfurt am Main, Germany}
\author{A. Koloczek}
\affiliation{Goethe-Universit\"at Frankfurt am Main, Germany}
\author{S.~Kr\"ackmann}
\affiliation{Goethe-Universit\"at Frankfurt am Main, Germany}
\author{C.~Langer}
\affiliation{Goethe-Universit\"at Frankfurt am Main, Germany}
\author{R. Plag}
\affiliation{Goethe-Universit\"at Frankfurt am Main, Germany}
\author{M. Pohl}
\affiliation{Goethe-Universit\"at Frankfurt am Main, Germany}
\author{G. Rastrepina}
\affiliation{Goethe-Universit\"at Frankfurt am Main, Germany}
\author{R. Reifarth}
\affiliation{Goethe-Universit\"at Frankfurt am Main, Germany}
\author{S. Schmidt}
\affiliation{Goethe-Universit\"at Frankfurt am Main, Germany}
\author{K. Sonnabend}
\affiliation{Goethe-Universit\"at Frankfurt am Main, Germany}
\author{M. Weigand}
\affiliation{Goethe-Universit\"at Frankfurt am Main, Germany}

\author{M.N.~Harakeh}
\affiliation{Kernfysisch Versneller Instituut, University of Groningen, Groningen, The Netherlands}
\author{N. Kalantar-Nayestanaki}
\affiliation{Kernfysisch Versneller Instituut, University of Groningen, Groningen, The Netherlands}
\author{C. Rigollet}
\affiliation{Kernfysisch Versneller Instituut, University of Groningen, Groningen, The Netherlands}
\author{S. Bagchi}
\affiliation{Kernfysisch Versneller Instituut, University of Groningen, Groningen, The Netherlands}
\author{M.A. Najafi}
\affiliation{Kernfysisch Versneller Instituut, University of Groningen, Groningen, The Netherlands}

\author{T. Aumann}
\affiliation{Technische Universi\"at Darmstadt, Germany}
\author{L.~Atar}
\affiliation{Technische Universi\"at Darmstadt, Germany}
\author{M. Heine}
\affiliation{Technische Universi\"at Darmstadt, Germany}
\author{M. Holl}
\affiliation{Technische Universi\"at Darmstadt, Germany}
\author{A. Movsesyan}
\affiliation{Technische Universi\"at Darmstadt, Germany}
\author{P. Schrock}
\affiliation{Technische Universi\"at Darmstadt, Germany}
\author{V. Volkov}
\affiliation{Technische Universi\"at Darmstadt, Germany}
\author{F. Wamers}
\affiliation{Technische Universi\"at Darmstadt, Germany}

\author{E. Fiori}
\affiliation{ExtreMe Matter Institute (EMMI), GSI D-64291 Darmstadt, Germany}
\affiliation{Frankfurt Institute for Advanced Studies FIAS,  Frankfurt am Main, Germany}
\author{B. L\"oher}
\affiliation{ExtreMe Matter Institute (EMMI), GSI D-64291 Darmstadt, Germany}
\affiliation{Frankfurt Institute for Advanced Studies FIAS,  Frankfurt am Main, Germany}
\author{J.~Marganiec}
\affiliation{ExtreMe Matter Institute (EMMI), GSI D-64291 Darmstadt, Germany}
\affiliation{Frankfurt Institute for Advanced Studies FIAS,  Frankfurt am Main, Germany}
\author{D. Savran}
\affiliation{ExtreMe Matter Institute (EMMI), GSI D-64291 Darmstadt, Germany}
\affiliation{Frankfurt Institute for Advanced Studies FIAS,  Frankfurt am Main, Germany}

\author{H.T. Johansson}
\affiliation{Chalmers Tekniska H\"ogskola, G\"oteborg, Sweden}

\author{P. Diaz Fern\'andez}
\affiliation{Universidade de Santiago de Compostella, Santiago de Compostela,
Spain}

\author{U. Garg}
\affiliation{University of Notre Dame, Department of Physics, Notre Dame, IN 46556} 
\affiliation{ExtreMe Matter Institute (EMMI), GSI D-64291 Darmstadt, Germany}

\author{D.L. Balabanski}
\affiliation{Inst. of Nucl. Res. and Nucl. Energy,
Bulgarian Acad. of Sci., Sofia, Bulgaria}

\begin{abstract}
The $\gamma$-decay of the anti-analog of the giant dipole resonance (AGDR) has been measured 
to the isobaric analog state  excited in the $p$($^{124}$Sn,$n$) reaction at a beam energy 
of 600 MeV/nucleon. 
The energy of the transition was also calculated with state-of-the-art self-consistent random-phase
approximation (RPA) and turned out to be very sensitive   to the neutron-skin thickness ($\Delta R_{pn}$).  By comparing the theoretical results with the measured one, the $\Delta R_{pn}$ value for $^{124}$Sn was deduced to be 0.175 $\pm$ 0.048 fm, which agrees well with
the previous results.  The energy of the AGDR measured previously for
$^{208}$Pb was also used to determine the $\Delta R_{pn}$ for $^{208}$Pb. In
this way a very precise $\Delta R_{pn}$ = 0.181 $\pm$ 0.031 neutron-skin
thickness has been obtained for $^{208}$Pb. The present method offers new
possibilities for measuring the neutron-skin thicknesses of very exotic isotopes.
\end{abstract}


\maketitle

Recent progress in development of radioactive beams has made it possible to
study the structure of nuclei far from stability. An important issue is the
size of the neutron skin of unstable neutron-rich nuclei, because this feature
 may provide fundamental nuclear structure information. There is a
renewed interest in  measuring precisely the thickness of the neutron skin,
because it  constrains the symmetry-energy term of the nuclear
equation of state. The precise knowledge of the symmetry energy is essential
not only for describing the structure of neutron-rich nuclei, but also for
describing the properties of the neutron-rich matter in nuclear astrophysics.

The  symmetry energy determines to a large extent, through the Equation of State
(EoS), the proton fraction of neutron stars \cite{la01}, the neutron skin in
heavy nuclei \cite{fu02} and enters as input in the analysis of heavy-ion
reactions \cite{li98, ba02}, etc.  Furnstahl \cite{fu02} demonstrated that in
heavy nuclei there exists an almost linear empirical correlation between the
neutron-skin thickness and theoretical predictions for the symmetry energy of
the EoS in terms of various mean-field approaches.  This observation has
contributed to a renewed interest in an accurate determination of the
neutron-skin thickness in neutron-rich nuclei \cite{te08,ta11,ro11,ab12}.
In this work, we are suggesting a new precise method for measuring the
neutron-skin thickness using both stable and radioactive beams.

In  our previous
work on inelastic alpha scattering, excitation of the isovector giant dipole
resonance was used to extract the neutron-skin thickness of nuclei \cite{kr91,
  kr94}. The cross section of this process depends strongly on $\Delta
R_{pn}$.  Another tool used earlier for studying the neutron-skin
thickness, is the excitation of the isovector spin giant dipole resonance
(IVSGDR). The L=1 strength of the IVSGDR is sensitive to the neutron-skin
thickness \cite{kr99, kr04}. 

 Vretenar
et al. \cite{vr03} suggested another new method for determining the $\Delta R_{pn}$
by measuring the energy of the GTR.  Constraints on the nuclear symmetry
energy and neutron skin were also obtained recently from studies of the
strength of the pygmy dipole resonance \cite{kl07}. In Ref. \cite{ca10},
more nuclei were added and the theory was better constrained.

The aim of the present work is to study the energy of the anti-analog of the
giant dipole resonance (AGDR) \cite{st80}, which depends strongly on the
neutron-skin thickness.   
The non-energy-weighted sum rule (NEWSR) we used earlier \cite{kr99, kr04} is valid (apart from a factor
of 3) also for the giant dipole resonance excited in charge-exchange reactions
and predicts increasing strengths  as
a function of the neutron-skin thickness.  Auerbach et al. \cite{au81} 
derived an energy-weighted sum rule (EWSR) also for the dipole strengths excited in
charge-exchange reactions. The
corresponding energies are measured with respect to the RPA g.s. energy (IAS
state)  in the parent. The result of such EWSR is almost
independent of the neutron-skin thickness \cite{au81} so the mean energy of
the dipole strength should decrease with increasing dipole strengths and 
therefore with increasing neutron-skin thickness in consequence of NEWSR. The
strong sensitivity of the AGDR energy on  $\Delta R_{pn}$ is mentioned also by
Krmpoti\'c \cite{kr83}, who preformed calculations with random-phase
approximation (RPA). In the present work, we want to use such sensitivity of
the energy of the AGDR on $\Delta R_{pn}$ to constrain the $\Delta R_{pn}$ of
$^{124}$Sn. 

Due to the isovector nature of the (p,n) reaction, the strength of the E1
excitation is divided into T$_0$-1, T$_0$ and T$_0$+1 components, where T$_0$ is
the ground state isospin of the initial nucleus, which is 12 for $^{124}$Sn. 
Because of the relevant  Clebsch-Gordan coefficients \cite{os92}, the T$_0$-1 component 
(AGDR) is favored compared to the T$_0$ and T$_0$+1 one by about a
factor of T$_0$, and 2T$_0^2$, respectively.  

Dipole resonances were excited earlier in ($p$,$n$) reactions at E$_p$= 45 MeV
by Sterrenburg et al. \cite{st80} in $^{92}$Zr, $^{93}$Nb, $^{94}$Mo,
$^{120}$Sn and in $^{208}$Pb.  Nishihara et al. \cite{ni85} measured also the
dipole strength distributions at E$_p$ = 41 MeV.  However, it was shown
experimentally \cite{os81,au01} that the observed $\Delta L$= 1 resonance was
in general a superposition of all possible spin-flip dipole (SDR) modes and
the non-spin-flip dipole
GDR.  According to the work of Osterfeld \cite{os92}
the spin non-flip/spin-flip ratio is favored at low bombarding energy (below
50 MeV) and also at very high bombarding energies (above 600 MeV).

The experiments, aiming at studying the neutron-skin thickness of $^{124}$Sn,
were performed at GSI using 600 MeV/nucleon $^{124}$Sn relativistic heavy-ion
beams on 2 and 5 mm thick CH$_2$ and 2 mm thick C targets. 
This allowed us to subtract the contribution of the C to the yield measured from
the CH$_2$ target during the analysis.
(Doing the experiment at 600 MeV/nucleon, in iverse kinematics we could
increase the  target thickness by a factor of 40 compared to 50 MeV/nucleon
case without loosing the energy resolution.) 

According to the previous experimental studies \cite{st80, ni85} the
excitation energy of the AGDR is expected to be at E$_x$= 26 MeV.  The
differential cross section of the AGDR excited in ($p$,$n$) reaction was
calculated at E$_p$ = 600 MeV with the computer code DW81 \cite{ra67,sc81}.
The wave functions used by the code were constructed using the normal-mode
formalism \cite{ho95} with the code NORMOD \cite{normod}. The optical model
parameters were taken from Ref. \cite{lo80}. According to such calculations
the dipole cross section peaks at $\Theta_{CM} = 3^\circ$.

The E$_x$= 26
MeV and $\Theta_{CM} =3^\circ$ correspond in inverse kinematics in the
laboratory system to a scattered neutron with energy 
of E$_n\approx$ 2.4 MeV and $\Theta_{LAB} \approx 68^\circ$,
respectively.

The ejected neutrons were detected by a low-energy neutron-array (LENA)
ToF spectrometer \cite{la11}, which was developed in Debrecen and which was
placed at 1 m from the target and covered a laboratory scattering-angle region
of 65$^\circ\leq\Theta_{LAB}\leq75^{\circ}$. Similar neutron spectrometers
have been built also by Beyer et al. \cite{be07} and by Perdikakis et
al. \cite{pe09} and one of those was used recently as an effective tool for studying
Gamow-Teller giant resonances in radioactive nuclear beams \cite{sa11}.  

The
energy of  de-exciting $\gamma$-transitions was measured by six large cylindrical
($3.5''\times 8''$)  state-of-the-art
LaBr$_3$ $\gamma$-detectors placed at 31 cm from the target and $\Theta_{LAB}$= 21$^\circ$ in order to
use the advantage of the large Lorentz-boost. The large Doppler shift
($E_\gamma /E_0 = 2.33$) was taken into account in the analysis. The precise
energy and efficiency calibrations of the detectors were performed after the
experiments by using different radioactive sources and (p,$\gamma$) reactions
on different targets \cite{ci09, labr2012}.  
The response function of the detector was also checked up to 17.6 MeV and
could be reproduced well with GEANT Monte-Carlo simulations. In order to make
a correct 
energy calibration for the AGDR, the simulations
were extended up to 40 MeV and convoluted with a Gaussian function 
with the  width of the resonance. This
convolution 
caused about 10\% lowering of the positions of the peaks, which was then taken 
into account in the calibration of the detectors.
The $\gamma$-ray energy spectrum
measured in coincidence with the low-energy neutrons is shown in Fig. 1.

\begin{figure}[bht]
\includegraphics[width=90mm]{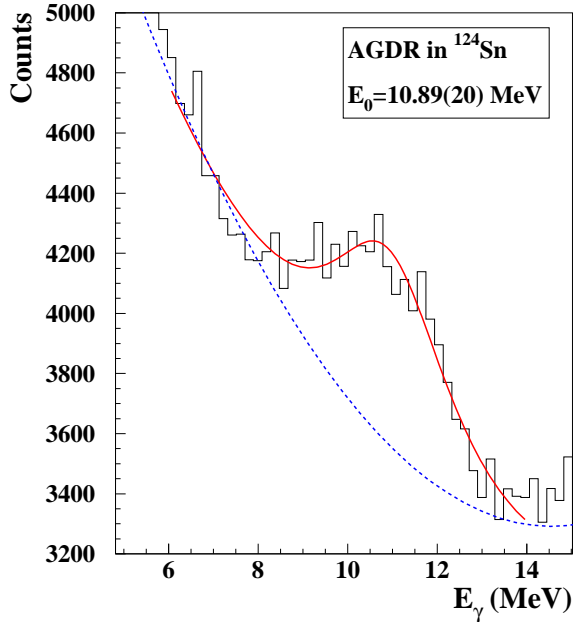}
\caption{The $\gamma$-ray energy spectrum measured in coincidence with the
  low-energy neutrons that fulfill the conditions of $1.0\leq E_n\leq 3.5$ MeV and
  67$^\circ \leq\Theta_{LAB}\leq 70^\circ$. The calibrated energy scale was corrected already for
  the Doppler effect. The solid line shows the result of the fit described in
  the text. 
\label{AGDR}}
\end{figure}

The width of the peak has a value of $\Gamma\approx$3.6 MeV, which is much
larger 
than the width of the previously measured AGDR resonances \cite{st80,ni85}. It can be explained by 
the Doppler broadening caused by the large solid angle of the detectors.
The energy distribution of the $\gamma$-rays was fitted by a Lorentzian curve and a
second order polynomial background, and the obtained parameters are shown in the figure. The
contribution of the statistical error in the uncertainty of the position of
the peak is  0.2 MeV, while the systematical error coming from the
uncertainty of the energy calibration is about 0.25 MeV ($2.5\%$), 
which can be improved in the future. If we
take into account the $E_\gamma^3$ dependence of the $\gamma$-transition
probability, then the $E_{AGDR}-E_{IAS}$ = 10.70 $\pm$ 0.32 MeV. 

The direct $\gamma$-branching
ratio of the AGDR to the IAS is expected to be similar to that of the GDR to
the g.s. 
in the parent nucleus, which can be
calculated from the parameters of the GDR  \cite{kr94}.
The branching ratio obtained for $^{124}$Sn is about 1\%. In contrast, in the 
investigation of the electromagnetic decay properties of the SDR
by Rodin and Dieperink \cite{ro02} the $\gamma$-decay branching ratio was in the range of
10$^{-4}$. This means that in the $n$-$\gamma$ coincidence spectrum the contribution of
the SDR is suppressed by about a factor of 100. Therefore, the coincidence
measurements deliver a precise energy for the AGDR, which in case of
$^{124}$Sn agrees well with the results obtained by Sterrenburg et
al. \cite{st80}.

The theoretical analysis employed in this work was carried out with 
the fully self-consistent relativistic
proton-neutron quasiparticle random-phase approximation (pn-RQRPA) based on
the Relativistic Hartree-Bogoliubov model (RHB) \cite{VALR.05}.  The RQRPA was
formulated in the canonical single-nucleon basis of the RHB model in
Ref.~\cite{Paar2003} and extended to the description of charge-exchange
excitations (pn-RQRPA) in Ref.~\cite{Paar2004}. The RHB + pn-RQRPA model is
fully self-consistent: in the particle-hole channel, effective Lagrangians
with density-dependent meson-nucleon couplings are employed, and pairing
correlations are described by the pairing part of the finite-range Gogny
interaction~\cite{BGG.91}.

For the purpose of the present study, we employ a family of density-dependent
meson-exchange (DD-ME) interactions, for which the constraint on the symmetry
energy at saturation density has been systematically varied, $a_4 =$ 30, 32,
34, 36 and 38 MeV, and the model parameters are adjusted to accurately
reproduce nuclear matter properties (the binding energy, the saturation
density, the compression modulus) and the binding energies and charge radii of
a standard set of spherical nuclei \cite{VNR.03}.  These effective
interactions were used to provide a microscopic estimate of the nuclear matter
compressibility and symmetry energy in relativistic mean-field
models~\cite{VNR.03} and in Ref.~\cite{kl07} to study a possible correlation
between the observed pygmy dipole strength (PDS) in $^{130,132}$Sn and the
corresponding values for the neutron-skin thickness. In addition to the set of
effective interactions with $K_{\rm nm} =$ 250 MeV (this value reproduces the excitation
energies of giant monopole resonances), and $a_4 =$ 30, 32, 34, 36 and 38 MeV,
the relativistic functional DD-ME2 \cite{LNVR.05} will be used here to
calculate the excitation energies of the AGDR with respect to the 
IAS, as a function of the neutron skin.
Important for the present analysis is the fact that the relativistic RPA
with the DD-ME2 effective interaction predicts the dipole polarizability
for $^{208}$Pb, $\alpha_D$=20.8 fm$^3$, in very good
agreement with the recently measured value: $\alpha_D = (20.1\pm 0.6)$
fm$^3$ \cite{ta11}.


The results of the calculations for $^{124}$Sn are shown in Fig. 2. The
difference in the excitation energy of the AGDR and the IAS, calculated with
the pn-RQRPA based on the RHB self-consistent solution for the ground-state of
the target nucleus, is plotted as a function of the corresponding RHB
prediction for the neutron-skin thickness.  For the excitation energy of the
AGDR we take the centroid of the theoretical strength distribution, calculated
in the energy interval above the IAS that corresponds to the measured spectrum
of $\gamma$-ray energies: 6 to 14.8 MeV (cf. Fig. 2). A single peak is
calculated for the IAS. For effective interactions with increasing value of
the symmetry energy at saturation $a_4 =$ 30, 32, 34, 36 and 38 MeV (and
correspondingly the slope of the symmetry energy at saturation
\cite{VNPM.12}), we find an almost perfect linear decrease of $E(AGDR) -
E(IAS)$ with the increase of the neutron skin $\Delta R_{np}$.  The value
calculated with DD-ME2 ($a_4 =32.3 $MeV) is denoted by the star symbol. 

The uncertainty of the theoretical predictions for the neutron-skin thicknesses 
is estimated to be 10 \%. Such an uncertainty
was used earlier for the differences between the neutron and proton
radii for the nuclei $^{116}$Sn, $^{124}$Sn, and $^{208}$Pb in adjusting the
parameters of the effective interactions \cite{VNR.03,LNVR.05}. 
These effective interactions were also used to calculate
the electric dipole polarizability and neutron-skin thickness of $^{208}$Pb,
$^{132}$Sn and $^{48}$Ca, in comparison to the predictions of more than 40 non-relativistic and relativistic mean-field effective interactions \cite{Pie12}.
From the results presented in that work one can also assess the
accuracy of the present calculations.

 In
comparison to the experimental result for $E(AGDR) - E(IAS)$ we deduce the
value of the neutron skin thickness in $^{124}$Sn: $\Delta R_{np} = 0.175 \pm 0.048 $
fm (including the 10\% theoretical uncertainty). 
In Table I, this value is compared to previous results obtained with a
variety of experimental methods. Very good agreement has been obtained with the previous data, which supports the
reliability of our method.

\begin{figure}[ht]
\includegraphics[width=90mm]{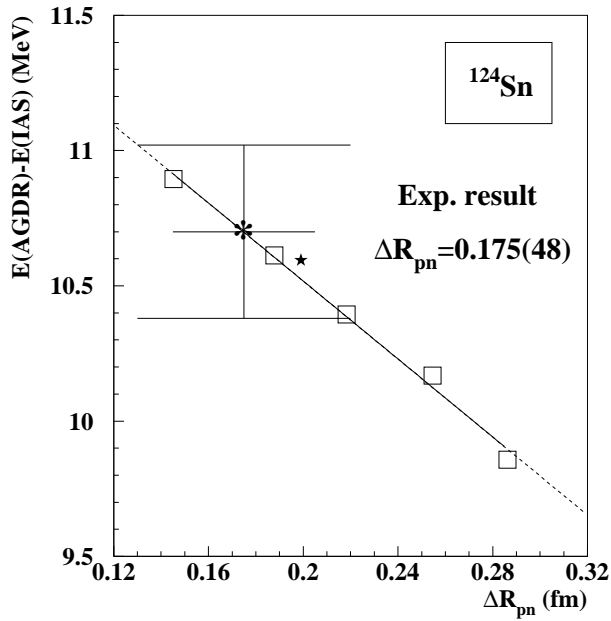}
\caption{The difference 
in the excitation energy of the AGDR and the IAS for the 
target nucleus $^{124}$Sn, calculated with the pn-RQRPA 
using five relativistic effective interactions characterized 
by the symmetry energy at saturation 
$a_4 =$ 30, 32, 34, 36 and 38 MeV (squares), and the interaction 
DD-ME2 ($a_4 =32.3$ MeV) (star). The theoretical values $E(AGDR) - E(IAS)$ 
are plotted as a function of the corresponding ground-state neutron-skin 
thickness $\Delta R_{pn}$, and compared to the experimental 
value $E(AGDR) - E(IAS) = 10.70 \pm 0.32 $ MeV. 
\label{skin124}}
\end{figure}

\begin{table}[ht]
\caption{\label{tab:table1} Neutron-skin thicknesses ($\Delta R_{pn}$) of
  $^{124}$Sn determined in the present work compared to previously measured
  values.  }
\begin{ruledtabular}
\begin{tabular}{lllc}
\textrm{Method}& \textrm{Ref.}& \textrm{Date}& \textrm{$\Delta R_{pn}$} (fm)\\ 
\colrule ($p$,$p$) 0.8 GeV & \cite{ra79,ba89} & 1979 & 0.25 $\pm$ 0.05 \\ 
($\alpha, \alpha$') GDR 120 MeV & \cite{kr94} & 1994 & 0.21 $\pm$ 0.11\\ 
($^3$He,$t$) SDR+GDR 177 MeV & \cite{kr04} & 2004 & 0.27 $\pm$ 0.07\\ 
antiproton absorption & \cite{tr01} & 2001 & 0.19 $\pm$ 0.02 \\ 
pygmy res.             & \cite{kl07} & 2007 & 0.24 $\pm$ 0.04 \\
($p$,$p$) 295 MeV & \cite{te08} & 2008 & 0.185 $\pm$ 0.017 \\ 
AGDR & pres. res. & 2012 &  0.175  $\pm$ 0.048 \\
\end{tabular}
\end{ruledtabular}
\end{table}

As the $n$-$\gamma$ coincidence method delivered similar results in the case of
$^{124}$Sn for the energy of the AGDR as obtained by Sterrenburg et
al. \cite{st80} by a ToF method, it is reasonable to assume that their result for the energy of
the AGDR obtained for $^{208}$Pb is correct.

Fig. 3 displays the corresponding theoretical results for $E(AGDR) - E(IAS)$ for 
the target nucleus $^{208}$Pb, as a function of the corresponding 
ground-state neutron-skin thickness $\Delta R_{pn}$, and compared 
with the experimental result of Sterrenburg et al. \cite{st80}. 
In this case the $E(AGDR)$ is calculated as the centroid of the theoretical 
strength distribution in the interval between 5 and 15 MeV above the IAS. 
The deduced value of the  neutron-skin thickness is compared with 
previous results in Table II.

\begin{figure}[htb]
\includegraphics[width=90mm]{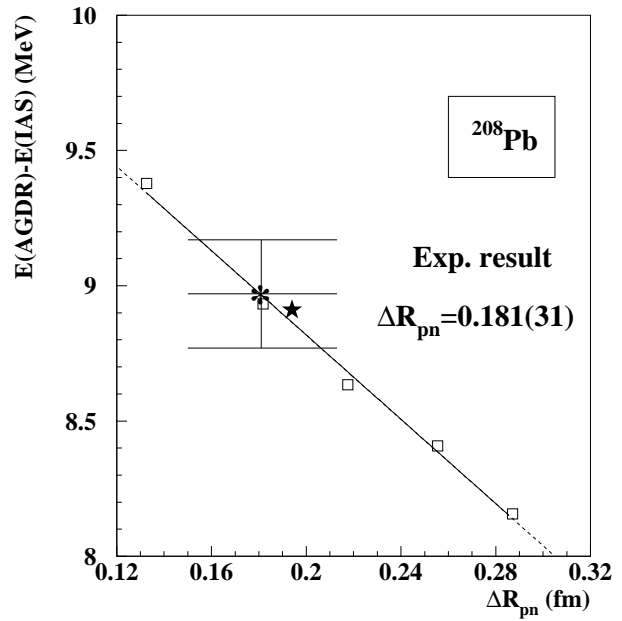}
\caption{Same as described in the caption to Fig. 2 but for the target nucleus $^{208}$Pb.
\label{Pb208}}
\end{figure}

\begin{table}[htb]
\caption{\label{tab:table2}Neutron-skin thicknesses of $^{208}$Pb determined
  in the present work compared to previously measured values.}
\begin{ruledtabular}
\begin{tabular}{lllc}
\textrm{Method}& \textrm{Ref.}& \textrm{Date}& \textrm{$\Delta  R_{pn}$} (fm)\\ 
\colrule ($p$,$p$) 0.8 GeV & \cite{ho80} & 1980 & 0.14 $\pm$ 0.04\\ 
($p$,$p$) 0.65 GeV & \cite{sta94} & 1994 & 0.20 $\pm$ 0.04 \\ 
($\alpha,\alpha$') GDR 120 MeV & \cite{kr94} & 1994 & 0.19 $\pm$ 0.09 \\ 
antiproton absorption & \cite{tr01} & 2001 & 0.18 $\pm$ 0.03 \\ 
($\alpha, \alpha$') GDR 200 MeV & \cite{kr04} & 2003 & 0.12 $\pm$ 0.07\\ 
pygmy res.             & \cite{kl07} & 2007 & 0.180 $\pm$ 0.035 \\
pygmy res.             & \cite{ca10} & 2010 & 0.194 $\pm$ 0.024 \\
($\vec p$,$\vec p\ '$) & \cite{ta11} & 2011 & 0.156 $\pm$ 0.025\\ 
parity viol. ($e$,$e$) & \cite{ab12} & 2012 & 0.33 $\pm$ 0.17 \\
AGDR & pres. res. & 2012 &  0.181 $\pm$ 0.031 \\
\end{tabular}
\end{ruledtabular}
\end{table}

%
In conclusion, we have investigated the energy of the AGDR excited in the
$^{124}$Sn($p$,$n$) reaction performed in inverse kinematics. Using the
experimental results from this study, Ref.~\cite{st80} for $^{208}$Pb, and
RHB+pn-RQRPA model, we deduce the following values of the neutron skin:
$\Delta R_{pn}$=(0.175 $\pm$ 0.048) fm in $^{124}$Sn and 0.181 $\pm$ 0.031
fm in $^{208}$Pb. The agreement between the $\Delta R_{pn}$ determined using
measurements of the AGDR-IAS and previous methods is very good in both the
studied cases.  In particular, the present study supports the results from
very recent high-resolution study of electric dipole polarizability $\alpha_D$
in $^{208}$Pb \cite{ta11}, respective correlation analysis of $\alpha_D$ and
$\Delta R_{pn}$ \cite{Pie12}, as well as the Pb Radius Experiment (PREX)
using parity-violating elastic electron scattering at JLAB \cite{ab12}. The
method we have introduced provides not only stringent constraint to the
neutron-skin thicknesses in nuclei under consideration, but it also offers new
possibilities for measuring $\Delta R_{pn}$ in rare-isotope beams.

\noindent{\bf Acknowledgments} This work has been supported by the European
Community FP7 - Capacities, contract ENSAR n$^\circ$ 262010, the Hungarian OTKA
Foundation No.\, K72566, the Croatian Science Foundation, the Bulgarian
Science Found, contacts DID-02/16 and DRNF-02/5, the U. S. National Science
Foundation (Grant No. 1068192), by the TÁMOP-4.2.2/B-10/1-2010-0024 project,
by the Alliance Program of the Helmholtz
Association (HA216/EMMI) and partially supported by the Italian Istituto 
Nazionale di Fisica Nucleare. UG acknowledges the support of EMMI during his
sojourn at GSI.


\begin{thebibliography}{99}

\bibitem{la01} J. M. Lattimer and M. Prakash, Astrophys. J. {\bf 550}, 426 (2001); astro-ph/0002232.  
\bibitem{fu02} R. J. Furnstahl, Nucl. Phys. {\bf A706}, 85 (2002).  
\bibitem{li98} B. A. Li, C. M. Ko, and W. Bauer, Int. J. Mod. Phys. {\bf E 7}, 147 (1998).  
\bibitem{ba02} Bao-An Li, Phys. Rev. Lett. {\bf 88}, 192701 (2002);  Nucl. Phys. {\bf A708}, 365 (2002). 
\bibitem{te08} S. Terashima et al., Phys. Rev. {\bf C 77},  024317 (2008).
\bibitem{ta11} A. Tamii et al., Phys. Rev. Lett. {\bf 107},  062502 (2011).
\bibitem{ro11} X. Roca-Maza and M. Centelles, X. Vinas, M. Warda, Phys. Rev. Lett. {\bf 106}, 252501 (2011). 
\bibitem{ab12} S. Abrahamyan et al., Phys. Rev. Lett. {\bf 108}, 112502 (2012). 
\bibitem{kr91} A. Krasznahorkay et al., Phys. Rev. Lett. {\bf 66}, 1287 (1991).
\bibitem{kr94} A. Krasznahorkay  et al., Nucl. Phys.  {\bf A567},   521 (1994). 
\bibitem{kr99} A. Krasznahorkay  et al., Phys. Rev. Lett. {\bf 82}, 3216 (1999).
\bibitem{kr04} A. Krasznahorkay et al., Nucl. Phys. {\bf A731}, 224 (2004).
\bibitem{vr03} D. Vretenar, N. Paar, T. Niksic, P. Ring, Phys. Rev. Lett.  {\bf 91},  262502 (2003).
\bibitem{kl07} A. Klimkiewicz et al., Phys. Rev. {\bf C 76},  051603 (2007).
\bibitem{ca10} A. Carbone et al., Phys. Rev. {\bf C 81},  041301 (2010).
\bibitem{st80} W.A. Sterrenburg, S.M. Austin, R.P. DeVito, A. Galonsky, Phys. Rev. Lett.  {\bf 45},  1839
  (1980).
\bibitem{au81} N. Auerbach, A. Klein, and Nguyen van Gai, Phys. Lett. {\bf B
  106} 347 (1981).

\bibitem{kr83} F. Krmpotic et al., Nucl. Phys.  {\bf A399},  478 (1983).
\bibitem{os92} F. Osterfeld,  Rev. Mod. Phys. {\bf 64}, 491 (1992).
\bibitem{ni85} S. Nishihara et al.,  Phys. Lett. {\bf B 160}, 369 (1985).
\bibitem{os81} F. Osterfeld et al., Phys. Lett. {\bf B 105}, 257 (1981).
\bibitem{au01} S. M. Austin et al.,  Phys. Rev. {\bf C 63}, 034322 (2001).
\bibitem{ra67} J. Raynal, Nucl. Phys. {\bf A97}, 572 (1967).
\bibitem{sc81} R.  Schaeffer and J.  Raynal, computer code  DWBA70 (1971), 
unpublished, extended version DW81 by J.R. Comfort (1981).
\bibitem{ho95} M. Hofstee et al., Nucl. Phys. {\bf A588},  729 (1995).
\bibitem{normod} S.Y. van der Werf, Program NORMOD, unpublished.
\bibitem{lo80} W.G.  Love and M.A.  Franey, Bull. Am. Phys.  Soc. {\bf
25} (1980) 730; W.G. Love private communication.
\bibitem{la11} C. Langer et al., Nucl. Instr. Meth. Phys. Res. {\bf A659},  411  (2011). 
\bibitem{be07} R. Beyer et al., Nucl. Instr. Meth. Phys. Res. {\bf A575},  449 (2007).
\bibitem{pe09} G. Perdikakis et al., IEEE Trans. Nucl. Sci. {\bf 56},  1174 (2009).
\bibitem{sa11} M. Sasano et al., Phys. Rev. Lett. {\bf 107}, 202501 (2011). 
\bibitem{ci09} M. Ciemala et al.,  Nucl. Instr.  Meth. Phys. Res. {\bf A608}, 76 (2009).
\bibitem{labr2012} A. Giaz et al., to be submitted to  Nucl. Instr. Meth. Phys. Res.
\bibitem{ro02} V. A. Rodin and A. E. L. Dieperink, Phys. Lett. {\bf B 541}, 7  (2002).
\bibitem {VALR.05} D. Vretenar, A. V. Afanasjev, G. A. Lalazissis, and
  P. Ring, Phys. Rep. {\bf 409}, 101 (2005).
\bibitem{Paar2003} N. Paar, P. Ring, T. Nik\v si\' c, and D. Vretenar,
  Phys. Rev. {\bf C 67}, 034312 (2003).
\bibitem{Paar2004} N. Paar, T. Nik{\v{s}}i{\'{c}}, D. Vretenar, and P. Ring,
  Phys. Rev. {\bf C 69}, 054303 (2004).
\bibitem{BGG.91} J. F. Berger, M. Girod, and D. Gogny, Comp. Phys. Comm.{\bf 63}, 365 (1991).
\bibitem{VNR.03} D. Vretenar, T. Nik{\v{s}}i{\'{c}}, and P. Ring, Phys. Rev.{\bf C 68}, 024310 (2003).
\bibitem{LNVR.05} G. A. Lalazissis, T. Nik{\v{s}}i{\'{c}}, D. Vretenar, and
  P. Ring, Phys. Rev. {\bf C 71}, 024312 (2005).
\bibitem{VNPM.12} D. Vretenar, Y.F. Niu, N. Paar, J. Meng, Phys. Rev.{\bf C 85}, 044317 (2012)
\bibitem{ra79} L. Ray, Phys. Rev. {\bf C 19}, 1855 (1979).
\bibitem{ba89} C. J. Batty et al., Adv. Nucl. Phys. {\bf 19}, 1 (1989).
\bibitem{tr01} A. Trzcinska et al., Phys. Rev. Lett.  {\bf 87}, 082501 (2001).
\bibitem{ho80} G.W. Hoffmann et al., Phys. Rev. {\bf C 21}, 1488 (1980).
\bibitem{sta94} V. E. Starodubsky and N. M. Hintz, Phys. Rev. {\bf C 49}, 2118 (1994).
\bibitem{Pie12} J. Piekarewicz et al., Phys. Rev.{\bf C 85}, 041302 (2012).


\end{thebibliography}
\end{document}